\documentclass[aps,11pt,showpacs,pre,reprint]{revtex4-1}
\usepackage{eurosym}    \usepackage{graphicx}  \usepackage{amssymb}
\usepackage{amsmath}
\begin{document}
\title{Characteristic distribution of finite-time Lyapunov exponents
  for  chimera states}  \author{A. E. Botha}  \email{Electronic
  address: {\em bothaae@unisa.ac.za} (A. E. Botha).}
\affiliation{Department of Physics, University of South Africa,
  Science Campus, Private Bag X6, Florida Park 1710, South Africa}
\date{\today }
\begin{abstract}
It is shown that probability densities of finite-time Lyapunov
exponents, corresponding to chimera states, have a characteristic
shape. Such distributions could be used as a signature of chimera
states, particularly in systems for which the phases  of all the
oscillators cannot be measured directly. In such cases, the
characteristic distribution may be obtained indirectly, via embedding
techniques, thus making it  possible to detect chimera states in
systems where they could  otherwise exist, unnoticed.   
\end{abstract}
    \maketitle
\section{Introduction}
Lyapunov characteristic exponents~\cite{shi79,ben80}, or more briefly
Lyapunov  exponents (LEs)~\cite{wol85} characterize the time-averaged
exponential divergence (positive exponents) or convergence (negative
exponents) of nearby orbits along orthogonal directions in the state
space. In numerical calculations of the LEs the asymptotic time
averaging is usually accomplished by using a sufficiently long time to
allow the averages of the exponents to converge within a set
tolerance. Although less frequently used, probability densities
(distributions) of the exponents, averaged over a much shorter time,
also contain valuable dynamic information. Such distributions are made
up of so-called finite-time, or local, Lyapunov exponents
(LLEs)~\cite{gra88,ott93,kap95}.  

For typical chaos the distribution of  LLEs can be accurately fitted
to a Gaussian function~\cite{gra88,ott93}, whereas for intermittent
chaos, at crises, and for fully developed chaos, the distributions are
characteristically non-Gaussian~\cite{pra99}. In the past the concept
of LLEs has been  used to characterize how secondary perturbations,
localized in space, grow and spread throughout distributed dynamical
(flow) systems with many degrees of freedom~\cite{pik93}. Variations
on this technique, i.e. of comoving or convective Lyapunov
exponents~\cite{pik93,rud96,gia00,men04}, continue to find new
applications in a variety of different contexts, ranging from
information theory~\cite{cen01,sch09} to fluid flow (see, for example,
Ref.~\cite{all15}, and the references therein). The notion of
finite-time Lyapunov exponents, averaged over initial conditions, has
also been  used to characterize transient chaos~\cite{ste10}.

In view of the fact that LLEs  have been employed successfully to
characterize many different types of nonlinear behaviour, it is
natural  to ask whether  a dynamical system in a so-called chimera
state~\cite{abr04}, may also possess a characteristic LLE
distribution. Chimera states are a relatively new type of
synchronization phenomenon. They occur in systems of (usually)
identical phase oscillators, which can be coupled,
nonlocally~\cite{kur02,abr04,abr06,pan13,sud15} (most frequently the
case), globally (all-to-all)~\cite{sch14} or even
locally~\cite{lai15}.  Depending on the nature  of the coupling  and
the initial conditions, the oscillators may divide up into two or more
spatially distinct groups, producing a spatiotemporal pattern which
simultaneously contains domains of coherent and incoherent
oscillations. However, such chimera states are fundamentally merely a
different type of deterministic (hyper)chaos, having one or more
positive LE(s)~\cite{wol11a,wol11b}. 

Although the existence  of chimera states was predicted more than a
decade ago in the seminal paper by Kuramoto and
Battogtokh~\cite{kur02}, experimental validation has only occurred
recently~\cite{abrphd,hag12,tin12,lar13,mar13,sch14,kap14,olm15,pan15b}.
Other than these fascinating experiments, chimera states may also be
of physical importance in systems of Josephson junctions
(JJs)~\cite{wie95,fil07a,fil07b}. Recently the spontaneous  appearance
of chimera states was found in  numerical simulations of so-called
SQUID metamaterials~\cite{laz15}. The superconducting quantum
interference devices (SQUIDs) are made of JJs. A SQUID metamaterial is
a one-dimensional linear array consisting of $N$ identical SQUIDs,
coupled together magnetically. The existence of a chimera state in
this model suggests that they may soon be detected experimentally in
existing one and two-dimensional SQUID metamaterials. At present there
is a renewed and ongoing interest in these intriguing materials, which
have even been proposed as a way of detecting quantum signatures of
chimera states~\cite{bas15}.

Certain highly anisotropic cuprate superconductors, such as
Bi$_2$Sr$_2$CaCu$_2$O$_{8+\delta}$, contain natural arrays of
intrinsic Josephson junctions (IJJs)~\cite{kle92}. At present there is
a concerted effort  being made towards achieving mutual
synchronization between stacks  of intrinsic JJs, with the view of
enhancing the power of the emitted  radiation in the terahertz
region~\cite{lin14}. In such systems the IJJs are  coupled together in
a way that is essentially nonlocal; a result of  the breakdown of
charge neutrality~\cite{ryn98}, or a diffusion
current~\cite{mac99,shu07}. IJJs could also provide a model for
studying other synchronization phenomenon, such as chaos
synchronization~\cite{sha15,bot15} and chimera states~\cite{bot16}.
To this end, one of the difficulties that must first be overcome is
related to the fact that, although  the voltage across a stack of
junctions can be measured with extreme precision, present experimental
setups do not provide direct access to the voltages across individual
junctions. Thus the states of the individual junctions have to be
inferred, somehow, from indirect measurements. This is where  the
customary method, of phase space reconstruction  via embedding and the
local function  approximation~\cite{wol85,ott93,all97,hil00} may play
an important role. In principle, the distribution of LLEs for a stack
of intrinsic JJs could be  obtained from a sufficiently long time
series of the total voltage across the stack; thus, making it possible
to detect the existence of a chimera state in the stack. At present
there are several highly sophisticated techniques that could
potentially be of used in this regard~\cite{tan96,cro10,yan11,yan12}.

While some previous studies of chimera states have computed their
LEs~\cite{wol11a,wol11b}, to the best of our knowledge, none has thus
far considered the distributions of LLEs for a chimera state. In view
of the above considerations, these distributions may in fact play an
important role in characterizing chimera states in general; but
particularly  in systems where the individual oscillators are not
accessible experimentally. To address this deficiency we compute
several distributions of LLEs corresponding to classic chimera states.
We show that these distributions have a common characteristic shape
that can be used to signal the occurrence of chimera states.  
   
\section{Model and methods}
As a basis for our investigation we consider a general class of
equations that support chimera states: 
\begin{eqnarray}
\frac{\partial \varphi(x_i,t)}{\partial t} & = & \omega_i -
\frac{K}{N}\sum_{j=1}^{N}C_{i j}G(x_i-x_j)\sin\left[\varphi(x_i,t)
  \right. \nonumber \\ & & \hspace*{2cm} -
  \left. \varphi(x_j,t)+\alpha\right] . \label{eq1}
\end{eqnarray}
A similar form to Eq.~(\ref{eq1}) was originally derived by Kuramoto
as an approximation to  the complex Ginzburg-Landau equation, under
weak coupling, when  amplitude changes may be neglected~\cite{kur84}.
With relatively few
exceptions~\cite{ome10b,hag12,tin12,mar13,ome13a,set13,bou14,gop14,hiz14,kap14,lik14,set14,sch14,zak14,laz15,pan15a},
the form of Eq.~(\ref{eq1}) encompasses the majority of systems that
have been considered in the literature on chimera states (see, for
example,
Refs.~\cite{kur02,abr04,abr06,bag07,abr08a,abr08b,set08,niy09,lai09,mar10,ome10a,hon11,wol11a,wol11b,ome12,zhu12,ome13b,pan13,yao13,xie14,xie15,pan15b}). It
describes the  dynamics of a non-locally coupled system of $N$ phase
oscillators, where $\varphi(x_i,t)$ is the phase of the $i$th
oscillator, located at position $x_i$. For identical oscillators the
distribution of natural frequencies is given by $\omega_i=\omega \;
\forall i$~\cite{kur02,abr04,pan13}. The function $G(x)$ has been
normalized to  have a unit integral and it describes the non-local
coupling between the oscillators~\cite{abr04}. $K$ controls the
overall coupling strength.  The coefficients $C_{ij}$ are either  a
coupling matrix consisting of ones and zeros, in the context of
networks~\cite{yao13},   or else they are quadrature weights, in
models  where large numbers of oscillators have been
considered~\cite{kur02,abr04,abr06}. In the latter models the
oscillators are assumed to be continuously distributed throughout a
one-dimensional spatial domain, leading to an integro-differential
equation of the form
\begin{equation}
\frac{\partial \varphi }{\partial t} = \omega -\int G\left(
x-x^{\prime }\right) \sin \left[ \varphi \left(x,t\right) -\varphi
  \left( x^{\prime },t\right) +\alpha \right] \mathrm{d}x^{\prime }.
\label{eq2}
\end{equation}

Since neither $C_{ij}$ nor $G(x_i-x_j)$ depend on the phases, the
Jacobian matrix of the system (\ref{eq1}) can be expressed
analytically as
\begin{eqnarray}
J_{ik} & = & - \frac{K}{N}\sum_{j=1}^{N}C_{ij}G(x_i-x_j)\cos
\left[\varphi(x_i,t) \right. \nonumber \\ & & \hspace*{1.5cm} \left. -
  \varphi(x_j,t) + \alpha \right]\left( \delta_{ik} - \delta_{jk}
\right). \label{eq3}
\end{eqnarray}
Equation (\ref{eq3}) allows us to compute the LLEs via the standard
algorithm (see Appendix~\ref{appA} for details), which uses
Gram-Schmidt orthonormalization to avoid numerical round off
errors~\cite{shi79,ben80,wol85}. In particular, we make use of the
Fortran implementation by Wolf, Swift and Swinney~\cite{wol85} for the
case when the system Jacobian is known analytically. Their code
required only minor modifications. Other than changing the system
equations, the code was modified in such a way that  it could be
called repeatedly over successive segments of the trajectory, at the
same time returning a time series of the trajectory, from which  the
averages $\left< \dot{\varphi}_{i} \right>$ could be computed.  To
integrate the system Eq.~(\ref{eq1}) and its linearization, we
employed  a fifth-order Runge-Kutta integration scheme, with fixed
time step. 

\section{Results and discussion}
\begin{figure}[tbh!]
\includegraphics[width=0.5\textwidth]{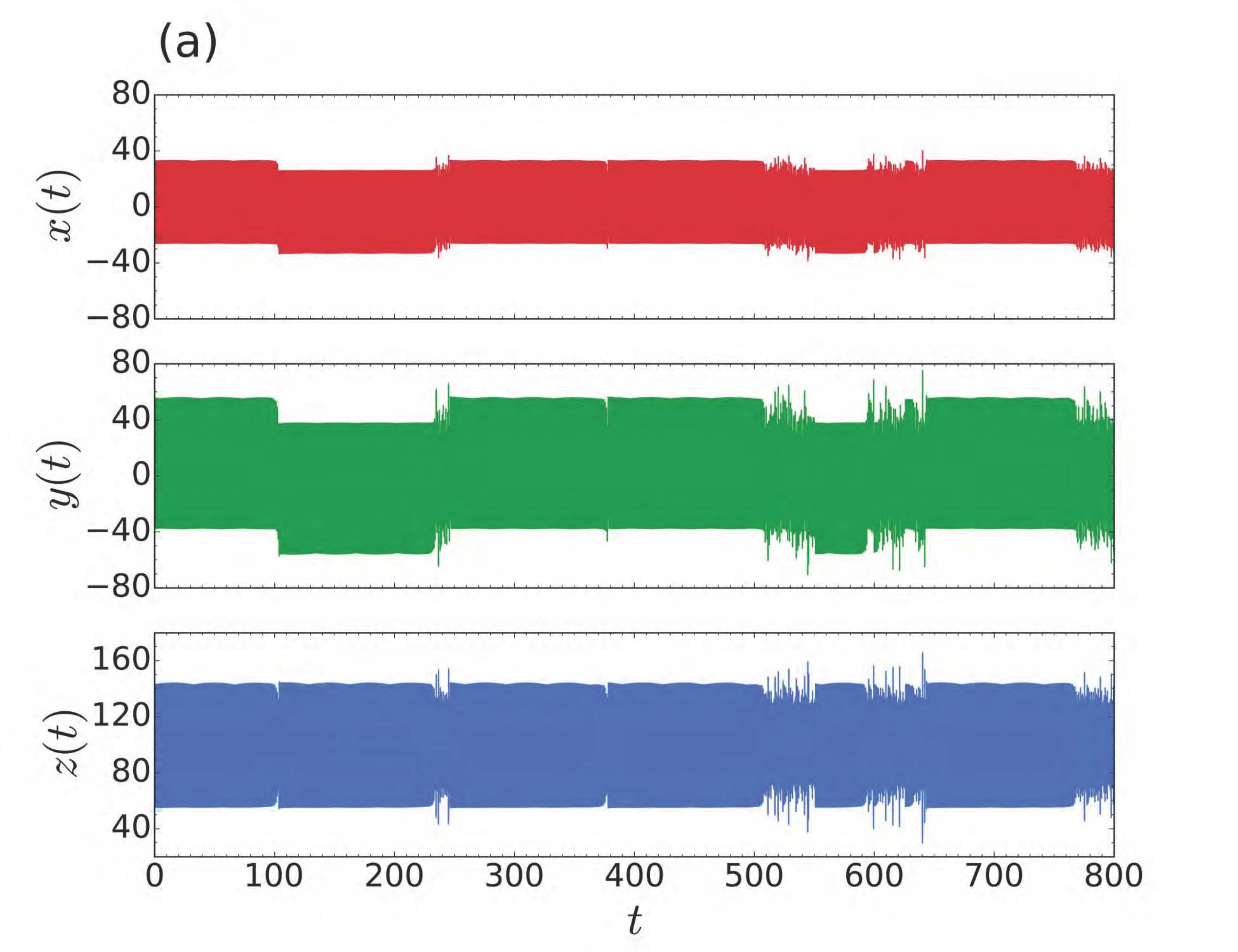}
\includegraphics[width=0.5\textwidth]{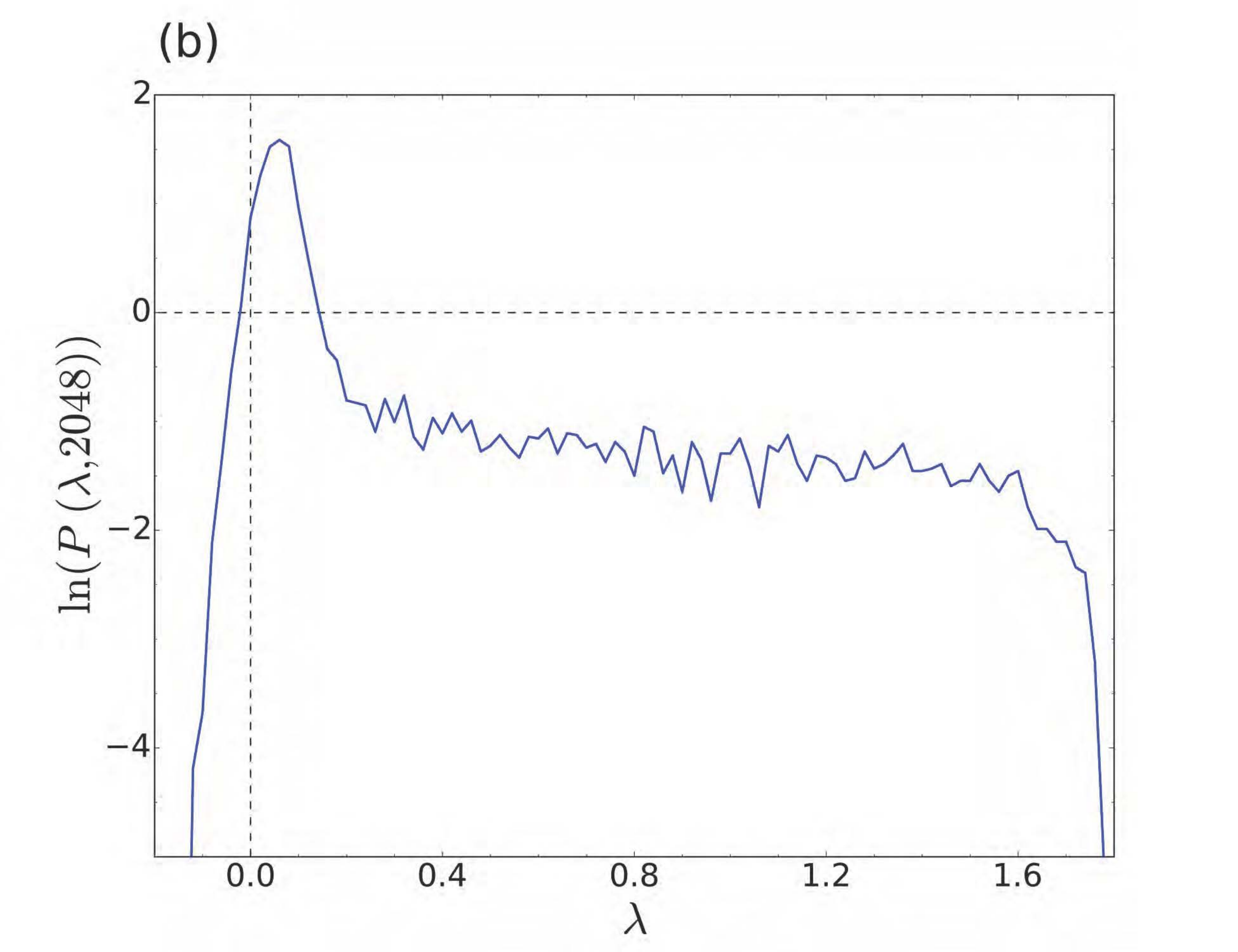}
\caption[bch!]{(Color online) (a) Time series showing intermittent
  chaos in the Lorenz system: $\dot{x} = a(y-x)$, $\dot{y} =
  x(r-z)-y$, $\dot{z} = xy -bz$, at parameter values $a=10$,
  $r=100.796$ and $b=8/3$. In this simulation a time step $\Delta t =
  1/64$ and initial condition (1,5,10) was used. (b) Logarithmic plot
  of the corresponding probability density, $P(\lambda,2048)$, for the
  distribution of ($10\,000$) maximal LLEs, $\lambda$, each  obtained
  by averaging over $2048$ time steps. The bin width was set to $0.02$
  and  the total simulation time was $320\,000$.}
\label{fig1}
\end{figure}

Before calculating the local Lyapunov exponent distributions for
chimera states, several test runs were made to reproduce known
distributions of LLEs, as reported in the
literature~\cite{kap95,pra99}. Fig.~\ref{fig1}, for example, 
shows the result of our calculations for the case of intermittent
chaos in the well-known Lorenz system, as discussed in
Ref.~\cite{pra99}.  As can be seen in Fig.~\ref{fig1}(a), the time
series for the system shows irregularly occurring  bursts of almost
periodic and chaotic behaviour. This motion corresponds to classic
(Type-I) tangent bifurcation intermittency~\cite{pom80}. In agreement
with  Fig.~6 of Ref.~\cite{pra99} the characteristic distribution of
LLEs consists of a superposition of two independent Gaussians, with
stretched exponential interpolation between the two. Qualitatively,
one  can rationalize the shape of the distribution by considering that
each Gaussian is roughly centered on the average value of the maximal
LEs that would characterize each type of motion separately, i.e. if
there was no switching. 

Although the characteristic distributions of LLEs are stationary over
a  wide range of averaging times, the averaging time used to compute
the LLEs does affect the widths of the distributions~\cite{pra99}. Our
numerical calculations confirm this observation. For very short
averaging times the distributions are not stationary, i.e. they keep
changing their shape, while in  the asymptotic  limit of infinite
averaging times, they all tend towards delta  functions. However,
provided these two extremes are avoided, the distributions are
stationary and maintain their characteristic shapes over a relatively
wide range of averaging times.

\begin{figure}[tbh!]
\includegraphics[width=0.5\textwidth]{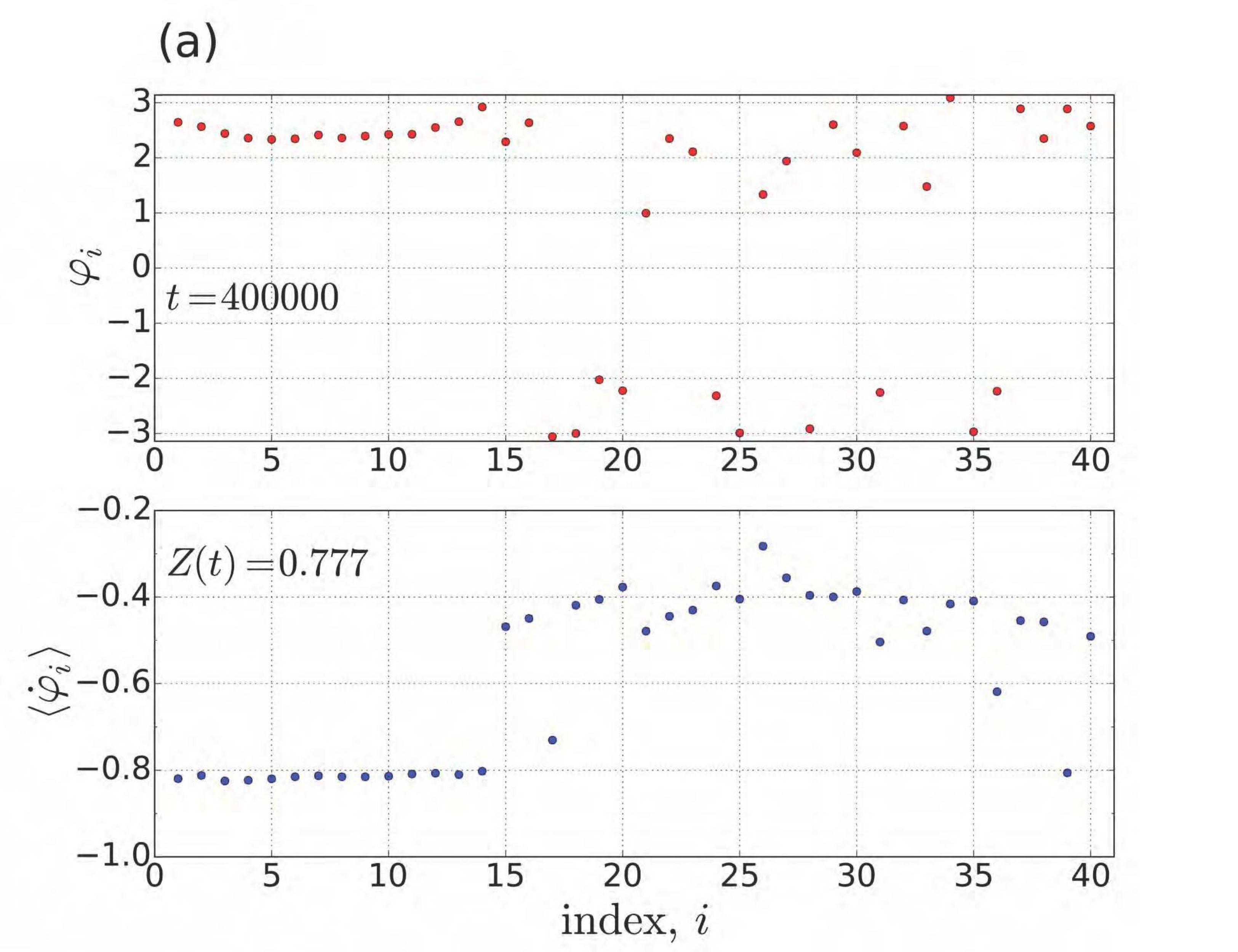}
\includegraphics[width=0.5\textwidth]{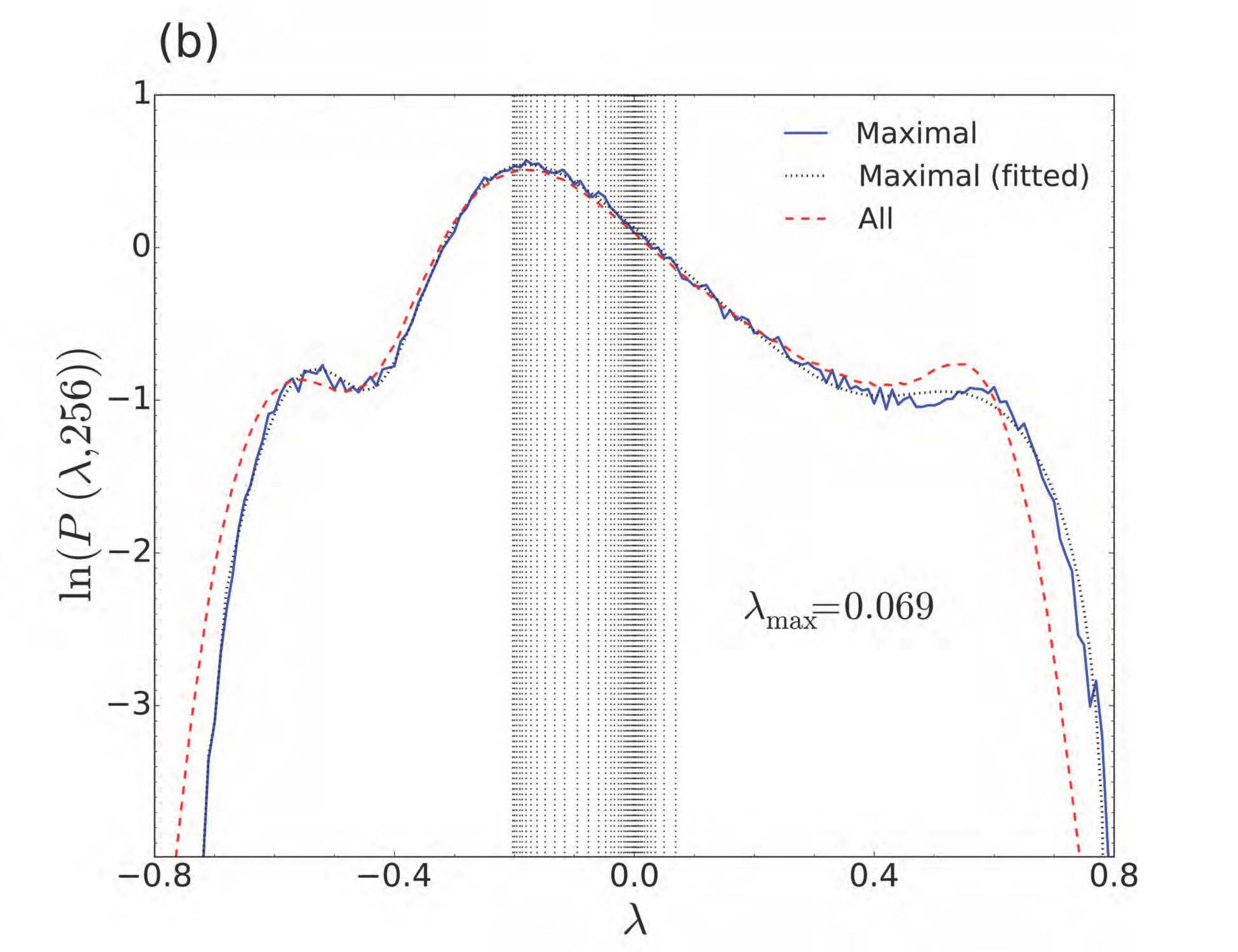}

\caption{(Color online) (a) A snapshot of the phases $\varphi_i$ and
  time averaged frequencies $\left< \dot{\varphi}_i \right>$ for
  system (1) in a chimera state. Parameters: $N = 40$, $R = 14$, and
  $\alpha = 1.478$. $\left< \dot{\varphi}_i \right>$ was averaged over
  $2048$ time steps, with $\Delta t = 1/128$.  (b) Corresponding
  probability distribution of the maximal LLEs (blue, solid line) and
  all LLEs (red, dashed line), averaged over $256$ time steps.  The
  vertical dotted lines indicate the values of the LEs, the largest
  being $\lambda_{\rm max} \! = 0.069$.}
\label{fig2}
\end{figure}

We now turn our attention to computing the LLEs for chimera states. We
begin by considering an interesting case of Eq.~(\ref{eq1}), that was
analysed in detail by Wolfrum  and Omel'chenko~\cite{wol11a}. The
parameters $N$ and $R$ in Ref.~\cite{wol11a} correspond to $K=N/(2R)$, 
\begin{equation}
C_{ij} = \left\{ \begin{array}{ll} 1 & \mbox{if }   | i - j| \le R
  \mbox{ or } | i - j| \ge N-R  \\ 0 & \mbox{otherwise,} \end{array}
\right. \label{eq4}
\end{equation}
and $G(x)=1$, in Eq.~(\ref{eq1}). In Fig.~\ref{fig2}(a) we show the
results of our simulation of this system  after $400\,000$ time
units. A snapshot of the distribution in  phases, $\varphi_i$, and
time averaged frequencies,  $\left< \dot{\varphi}_i\right>$ (as well
as the time series of  the order parameter $Z(t) =\left|
\sum_{k=1}^{N}\exp(i\varphi_k)\right| $, not shown), clearly indicate
that the oscillators are in a chimera state throughout the whole
simulation.  In Fig.~\ref{fig2}(b) the corresponding distributions for
the maximal LLE (blue, solid line) and all LLEs (red, dashed line),
both averaged over $256$ time steps, can be seen.                   
As observed in previous calculations of the LLEs~\cite{pra99}, the
distribution of all the exponents together has roughly the same shape
as that  of the maximal exponent alone. It is of course much smoother,
due to the larger number of samples in the distribution, and is
somewhat shifted towards the negative exponent side.     The
distribution shown in Fig.~\ref{fig2}(b) appears to be characteristic
to {\em all} chimera states arising from  the general class of
equations~(\ref{eq1}). It consists of a central asymmetric
Gaussian-like peak with shoulders on both sides. One notable common
feature in the distributions for chimera states  is their approximate
symmetry with respect to the $\lambda=0$ axis,  c.f. the intermittent
chaos distribution shown in Fig.~\ref{fig1}(b).  The feature of having
two shoulders, one to either side of the main central peak, appears to
be another distinguishing attribute specific to chimeras. To make
these qualitative observations more precise we have fitted a variety
of distributions corresponding to chimera states (all obtained from
Eq.~(\ref{eq1}) at a variety of (very) different parameters) and found
that all such distributions can be fitted accurately by a linear
combination of four fitting functions (see Appendix \ref{appB} for
details).  By contrast, distributions of LLEs for intermittent chaos
can be fitted to a comparable accuracy by a linear combination of only
two Guassians and one exponential function.
          
Based on their analysis, Wolfrum and Omel'chenko~\cite{wol11a}  have
suggested that, ``chimera states are chaotic transients''.
Notwithstanding  the experimental observations, there have been
several counter-examples  of numerically  simulated chimera states
that are stable, independent of the population  size or initial
conditions~\cite{sch14}. In a general sense, the claim  that chimera
states are chaotic transients is therefore certainly not 
valid~\cite{sud15}.
Surprisingly, however, there have been very few comments about why
this conclusion was reached for the prototype system~\cite{pan15b}.
In the course of performing the present simulations, we have observed
that the same system ($N=40$, $R=14$) is actually capable of
supporting a variety of different chimera states, depending on the
range of the ``phase lag'' parameter $\alpha$~\cite{pan13,abrphd}.  In
our view the existence of chimera states in this system can be
understood in terms of a balance between the tendency for the
oscillators to synchronize, and the tendency for them not to. At the
value $\alpha=1.46$,  considered in Ref.~\cite{wol11a}, the overall
coupling in the system was `attractive', thus causing the system to
synchronize after a certain time. The fact that the synchronization
time was shown to be  exponentially distributed (with respect to  $N$)
may be related to the probabilities of an individual oscillator to
have an almost matching phase with one or more of its neighboring
oscillators, i.e. as the system evolves. One can see this
qualitatively from the equations of motion, by considering how
$\alpha$ affects the slowing down or speeding up ($\dot{\varphi}$) of
any individual oscillator due to its coupling with an equidistant pair
of neighboring oscillators. To make these observations quantitative,
is beyond the scope of the present work. For the system  in
Ref.~\cite{wol11a} it suffices to say that, depending on the value of
$\alpha$, we have found that the  initial chimera state may (i)
synchronize completely after a certain time, (ii) persist (apparently)
indefinitely, (iii) only appear intermittently between bursts of
chaos, or (iv) collapse to an incoherent state after a very short
time, never to re-appear again. 

\begin{figure}[tbh!]
\includegraphics[width=0.5\textwidth]{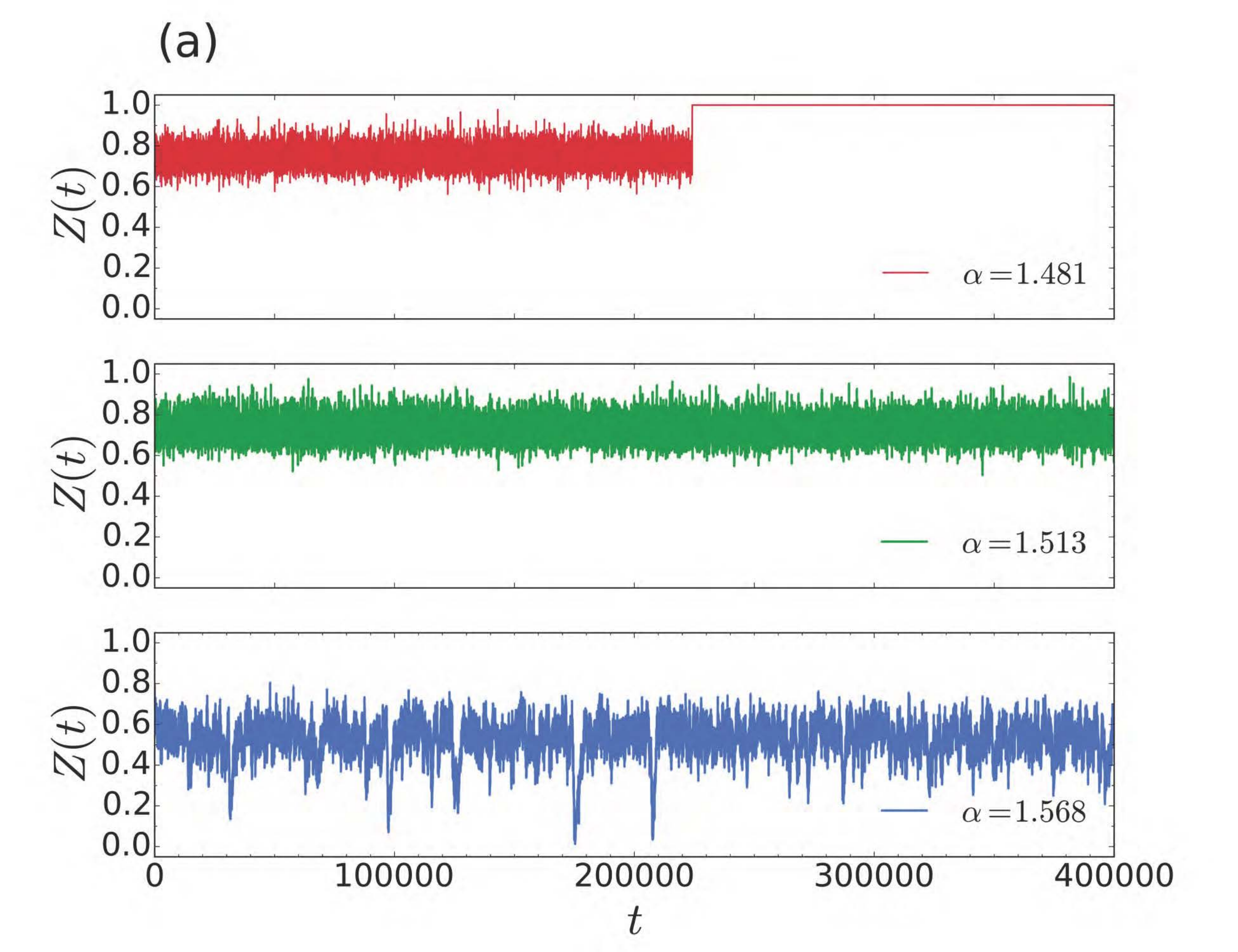}
\includegraphics[width=0.5\textwidth]{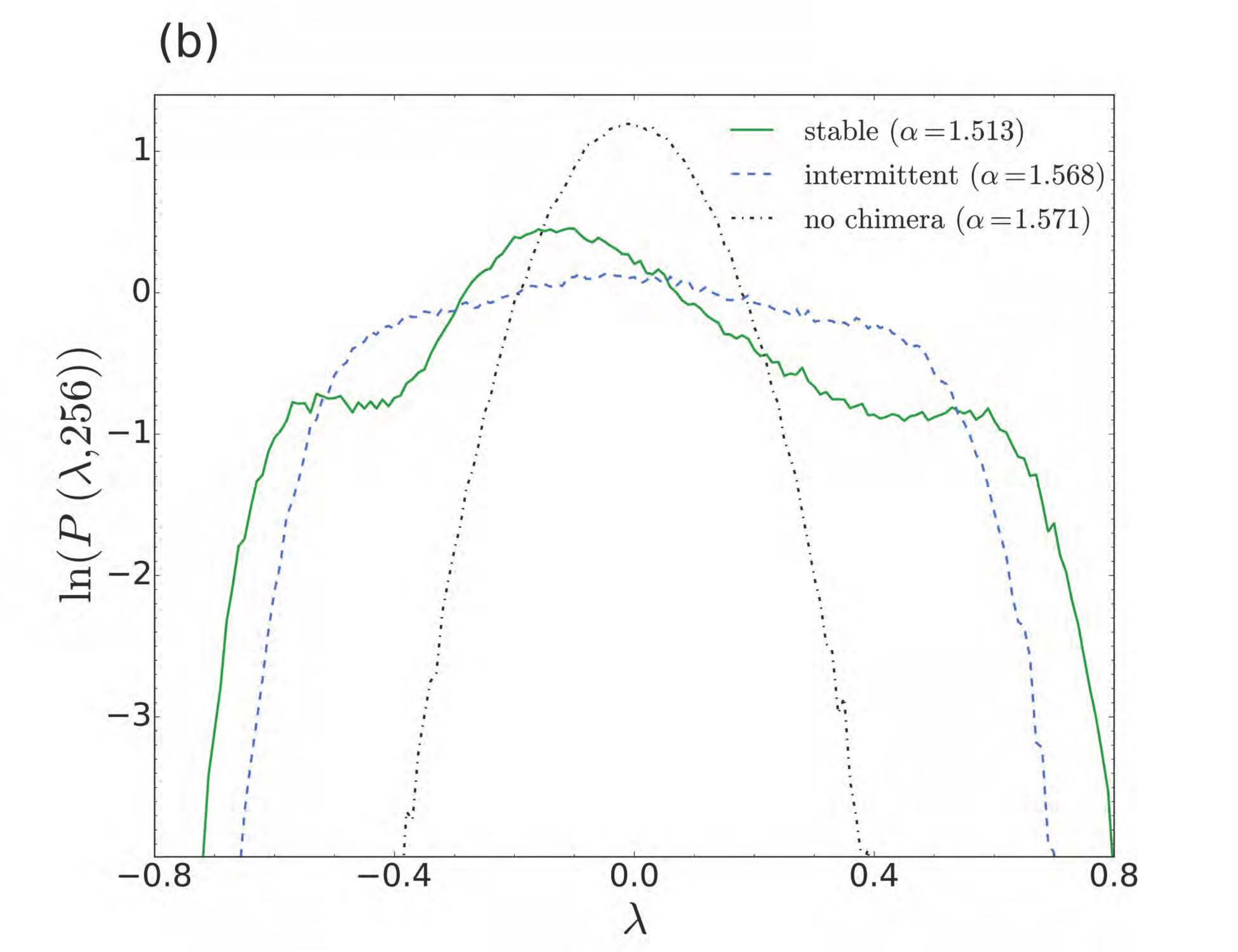}

\caption{(Color online) (a) Time series of the order parameter $Z(t)
  =\left| \sum_{k=1}^{N}\exp(i\varphi_k)\right| $ for three different
  types of chimera states. In the top figure ($\alpha = 1.481$) the
  state is transient and the oscillators all synchronize for $t
  \gtrapprox  225\,000$ In the middle figure ($\alpha=1.513$)  the
  state is stable, while in the bottom figure ($\alpha = 1.568$) it
  appears intermittently. (b) The characteristic distributions of the
  maximal LLEs for stable and intermittent chimera states. For
  comparison the Gaussian distribution, expected for typical chaos
  (with no chimera), is also shown.} \label{fig3}
\end{figure}

To investigate the effect of the above four scenarios on the
distribution  of LLEs, we have performed many simulations of the
($N=40$, $R=14$) system at different values of $\alpha$ in the range
$[1.460, 1.571]$. Each simulation  had a slightly perturbed initial
condition, as described in Ref.~\cite{wol11a}.  The outcomes of our
simulations all fall within the above categories: for $1.460 \leqslant
\alpha < 1.513$, not all the initial chimera states persisted to  the
end of the simulation time ($400\, 000$ time units).  For $1.513
\leqslant \alpha \leqslant 1.553$ all the chimera state persists
throughout the whole simulation time. For $ 1.553 < \alpha \leqslant
1.568$ chimera states appear intermittently, interrupted by bursts of
chaos. Lastly, for $1.568 < \alpha \leqslant 1.571$ the initial
chimera rapidly evolved into permanent incoherence.  In
Fig.~\ref{fig3}(a) we summarize these results by displaying time
series of the order parameters at three different values of $\alpha$.      
In the lower time series, for an intermittently occurring chimera at
$\alpha = 1.568$, two events can be seen where the order parameter
drops down close to zero, i.e. intermittently the system becomes
almost totally incoherent. For the majority of the simulation time
$Z(t)$ hovers around $0.6$. Random spot  checks on the phases and
averaged frequencies show that the system is  unambiguously in a
chimera state for times when $Z(t) \gtrapprox 0.4$.
Fig.~\ref{fig3}(b) shows the corresponding distributions of the
maximal  LLEs corresponding to the cases (ii), (iii) and (iv), as
described above. Here it can be seen that the characteristic
distribution for the chimera state abruptly becomes flat-topped when
the chimera starts to appear intermittently. Consistent with our
expectations, once the chimera disappears completely the distribution
corresponding to the incoherent oscillators is Gaussian in form. On
the logarithmic plot that is shown in Fig.~\ref{fig3}(b) it appears as
a parabola, corresponding to the case of typical chaos. 

We next consider the chimera state that was originally reported by
Abrams and Strogatz~\cite{abr04} for the system given by
Eq.~(\ref{eq1}), or (\ref{eq2}),  with the coupling
\begin{equation}
G(x) = \frac{1}{2\pi}\left(1 + 0.995\cos x \right) \mbox{.}
\label{eq5}
\end{equation}
As in Ref.~\cite{abr04}, we solve this system for $N=K=256$ and
$\alpha = \pi/2 - 0.18$, using Simpson's  $3/8$ rule~\cite{jef08}, for
which the  quadrature weights in Eq.~(\ref{eq1}) are given by: $C_{k\,
  1}=3h/8$, $C_{k\, 2}=9h/8$, $C_{k\, 3}=9h/8$, $C_{k\, 4}=6h/8$,
$C_{k\, 5}=9h/8$ , $C_{k\, 6}=9h/8$, $C_{k\, 7}=6h/8$, \ldots , $C_{k
  \, N-1}=9h/8$, $C_{k \, N}=3h/8$. Here the space variable $x$ runs
from $-\pi$ to $\pi$ with periodic boundary conditions, and $h=2\pi/N$
is the separation between the identical oscillators. 

\begin{figure}[tbh!]
\includegraphics[width=0.5\textwidth]{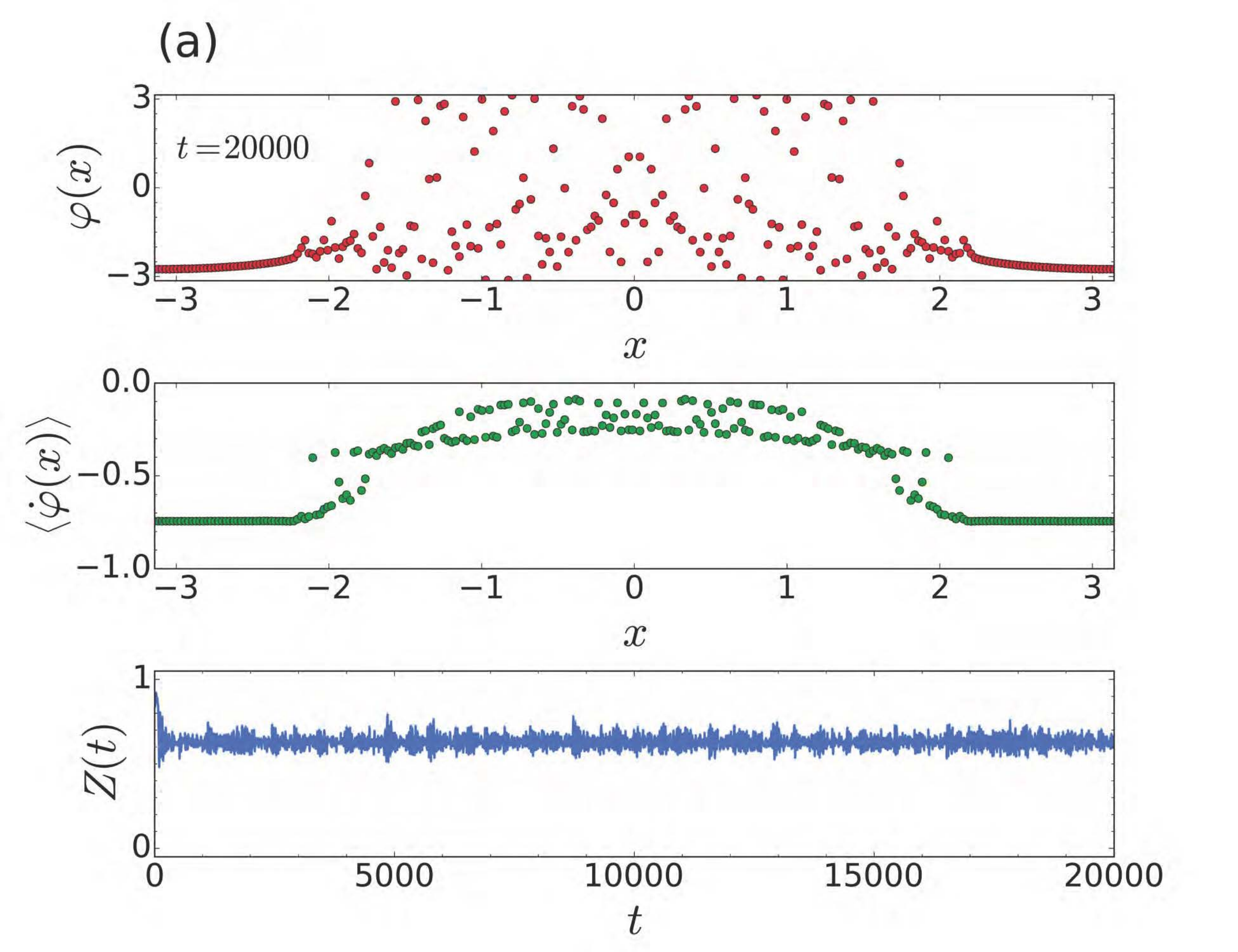}
\includegraphics[width=0.5\textwidth]{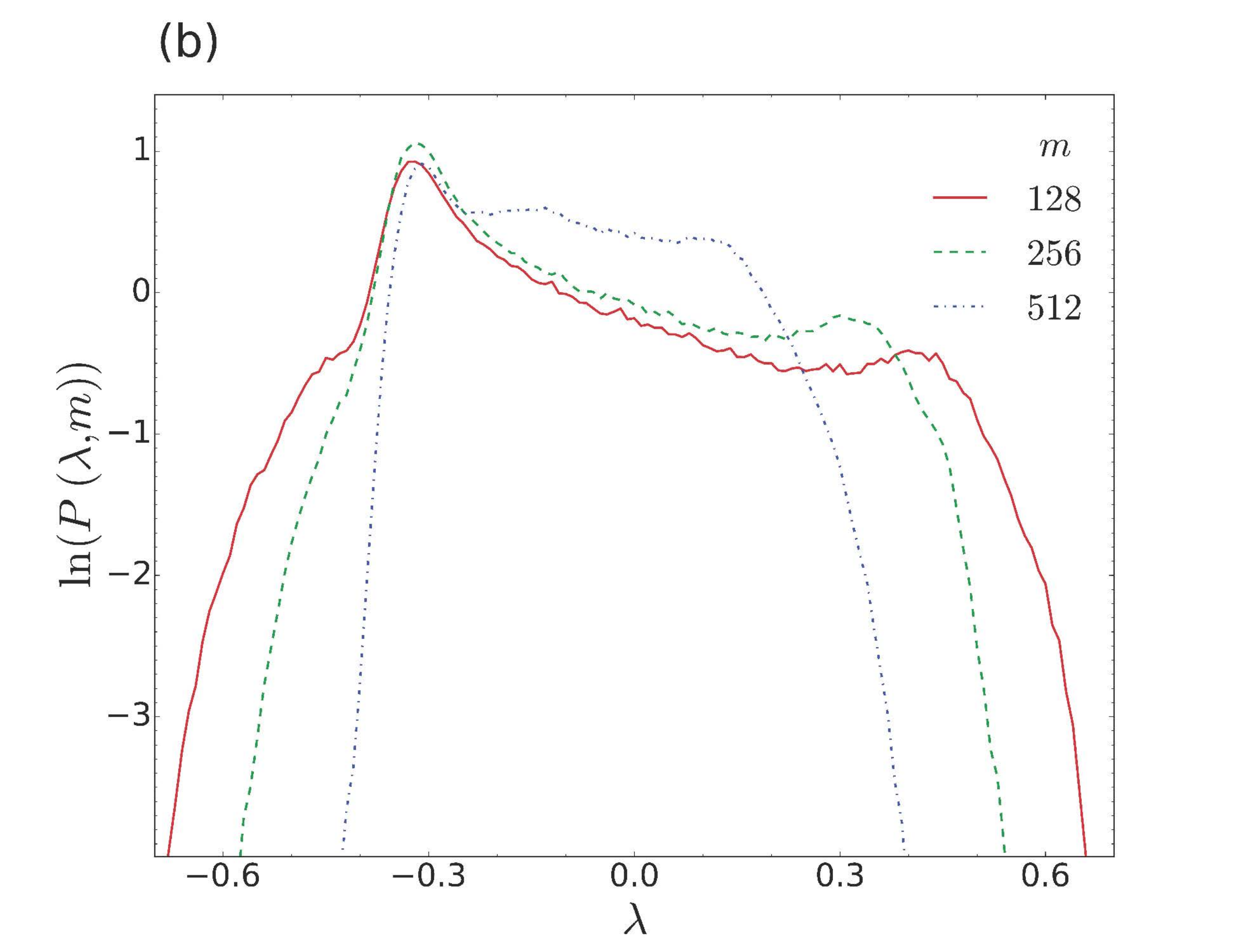}

\caption{(Color online) (a) The chimera state for the system described
  in Ref.~\cite{abr04}, showing a snapshot of the phases $\varphi(x)$
  and time averaged phase velocities $\left< \dot{\varphi}(x)\right>$,
  after an integration time of $20 000$ dimensionless units. The
  averaging for $\dot{\varphi}$ was done over $1024$ time steps, each
  of duration $\Delta t = 1/128$. The lower figure shows the time
  series of the order parameter $Z(t)$.  (b) Comparison of the
  characteristic distributions of all LLEs for the system in (a),
  obtained by averaging the LLEs over  $128$, $256$ and $512$ time
  steps, respectively.} \label{fig4}
\end{figure}

Fig.~\ref{fig4}(a) shows the distribution of phases,  averaged
frequencies, and the time series of the order parameter for this
chimera state. More importantly, Fig.~\ref{fig4}(b) shows the
distributions of all the LLEs, obtained by averaging over three
different time intervals. As was previously mentioned, although all
three distributions are stationary and maintain the  shape of the
characteristic distribution, their widths decrease as the averaging
time increases from $128$ to $512$ time steps.     

\section{Conclusion}
We have calculated the probability distributions of the  local
Lyapunov exponents (LLEs) corresponding to  chimera states and found
that they form a very characteristic distribution which appears to be
specific to the state's characteristic spatiotemporal pattern of
synergistic coherent and incoherent motion. In principle a knowledge
of this expected characteristic  distribution can be used to identify
the occurrence of chimera states in  real physical systems,
particularly in those for which it may not be possible  to measure,
directly the phases of of all the oscillators. In such cases, we
envisage that advanced embedding techniques could be employed to
extract  the characteristic distribution of LLEs, which would then be
a useful signature of the chimera state (for it would otherwise have
been undetectable). The present results may thus find application in a
variety systems where chimera states are relevant, but not necessarily
directly observable. A comprehensive review of the relevant real-world
systems may be found in Ref.~\cite{pan15b} and the references therein.
    
The case of intermittently appearing chimeras, as discussed in
connection with Fig.~\ref{fig3}, is currently of particular interest,
in view of the fact that  intermittent chaotic chimeras have only
recently been reported in the literature for a much more complicated
system~\cite{olm15}.  This system consisted of two symmetrically
coupled populations  of $N$ oscillators, where one population is
synchronized and the other  jumps erratically between laminar and
turbulent phases~\cite{olm15}.  The present considerations indicate
that such intermittency also arises in the original prototype system
for chimera states, although it was not previously reported. 

\begin{acknowledgments}
The author wishes to thank M. R. Kolahchi, W. Dednam and O. E. Omel'chenko for
helpful discussions about this work. He would also like to thank 
D. M. Abrams and S. H. Strogatz for their generous e-mail communications in 
answer to questions about some of their original simulation techniques.
\end{acknowledgments}

\appendix
\section{Probability density of LLEs defined} \label{appA}
Consider an autonomous $n$-dimensional continuous dynamical system of
the form $\boldsymbol{\dot{x}}=F\left( \boldsymbol{x} \right)
$. The probability density (distribution) of the $m$th LLE, 
$P\left( \lambda
^{m},N\right) $, is defined so that $P\left( \lambda
^{m},N\right) \mathrm{d} \lambda ^{m}$ equals the probability that
$\lambda _{N}^{m}$ takes on a value between $\lambda ^{m}$ and
$\lambda ^{m}+\mathrm{d}\lambda ^{m}$, where
\begin{equation}
\lambda _{N}^{m}=\frac{1}{N}\sum\limits_{j=1}^{N}\ln \left\Vert
\boldsymbol{ e}_{j}^{m}\right\Vert   \label{eqa1} \mbox{.}
\end{equation}
Here $\boldsymbol{e}_{0}^{m}$ is the $m$th vector of the initial set
of orthonormal vectors, and $\boldsymbol{e}_{j}^{m}$ is its time
evolved value under the action of the linearized equations of motion
after $j$ time steps. Thus the $\boldsymbol{e}_{j}^{m}$ are obtained
by solving the equations 
\begin{equation}
\dot{\boldsymbol{e}}^{m}= \left\{ \boldsymbol{J}F\left( \boldsymbol{x} \right) 
\right\} \boldsymbol{e}^{m}  \label{eqa2}
\mbox{,}
\end{equation}
where $\boldsymbol{J}$ is the system Jacobian. Notice that $\lambda
_{N}^{m}$ is the $m$th exponent averaged over $N$ time steps.  Because
the vectors $\boldsymbol{e}^{m}$ diverge exponentially in magnitude, and
tend to align themselves along the local direction of  most rapid
growth, their exact directions may rapidly become numerically
indistinguishable. To overcome this difficulty, as explained in
Refs.~\cite{shi79,ben80,wol85}, Gram-Schmidt orthogonalization can be
performed on the set of frame vectors. In the present work we have
performed the Gram-Schmidt orthogonalization after every time step.

\section{Fitting the characteristic distribution}  \label{appB}
It is found that the characteristic distribution of LLEs for chimera
states can be fitted accurately by a linear combination of the form
$g_1(x) + f(x) + g_2(x) + B\exp(-x/\tau)$, where $g_1$ and $g_2$ are
Gaussians given by 
\begin{equation}
g_i(x)= \frac{A_i}{\sigma_i\sqrt{2\pi}}\exp\left[-\left(
  \left(x-\mu_i\right)/\sqrt{2\sigma_i^2}
  \right)^2\right] \label{eqb1} \mbox{,} 
\end{equation} 
and $f(x)$ is an exponentially modified Gaussian, given by 
\begin{equation}
f(x) = \frac{A\gamma}{2}\exp\left( \frac{ \mu-x+
  \gamma\sigma^2}{2/\gamma} \right) \mbox{erfc} \left(
\frac{\mu+\gamma\sigma^2 - x}{\sqrt{2\sigma^2}}\right)\label{eqb2}
\mbox{,}
\end{equation}
where erfc is the complementary error function~\cite{jef08}. To
perform the fitting we made use of the Python  package
\verb|lmfit|~\cite{new14}, which stands for Non-Linear Least-Squares
Minimization and Curve-Fitting for Python. The initial parameters were
chosen so that the  Guassians were roughly centered on the two
shoulders of the distribution, with $f(x)$ on the central main
peak. In all cases, the distributions could be fitted with reduced
chi-square values  of less than 0.002. 

For the data corresponding to the solid (blue) curve in Fig.~\ref{fig2}(b), 
for example, the fitting routine produced the following output:
\begin{verbatim}
[[Model]]
    (((Model(gaussian, prefix='g1_') 
     + Model(expgaussian, prefix='f_')) 
     + Model(gaussian, prefix='g2_')) 
     + Model(exponential, prefix='exp_'))
[[Fit Statistics]]
    # function evals   = 970
    # data points      = 161
    # variables        = 12
    chi-square         = 0.088
    reduced chi-square = 0.001
[[Variables]]
    exp_amplitude:  -0.00123468 (init = 0.0)
    exp_decay:      -0.16029312 (init =-0.33)
    g1_sigma:        0.08416450 (init = 0.085)
    g1_center:      -0.53906326 (init =-0.54)
    g1_amplitude:    0.09116373 (init = 0.1)
    f_amplitude:     0.87642684 (init = 0.9)
    f_sigma:         0.09597549 (init = 0.09)
    f_center:       -0.29751003 (init =-0.3)
    f_gamma:         3.40994342 (init = 1)
    g2_sigma:        0.14054123 (init = 0.094)
    g2_center:       0.59176841 (init = 0.56)
    g2_amplitude:    0.09098822 (init = 0.054)
\end{verbatim}
The fitted distribution has also been plotted in Fig.~\ref{fig2}(b) 
as a dotted (black) line.

\end{document}